\documentclass[3p,12pt]{elsarticle}

\usepackage{amssymb}
\usepackage{amsmath}
\usepackage{amsfonts}
\usepackage{multirow}

\journal{Journal of Parallel and Distributed Computing}

\begin{document}

\begin{frontmatter}

\title{Highly scalable numerical simulation of coupled reaction-diffusion systems with moving interfaces}

\author[1]{Mojtaba Barzegari\corref{cor1}}
\ead{mojtaba.barzegari@kuleuven.be}

\author[1,2]{Liesbet Geris}
\ead{liesbet.geris@kuleuven.be}
\ead{liesbet.geris@uliege.be}
\ead[url]{www.biomech.ulg.ac.be}

\cortext[cor1]{Corresponding author, Tel: (+32) 16 193831}

\address[1]{Biomechanics Section, Department of Mechanical Engineering, KU Leuven, Leuven, Belgium}
\address[2]{Biomechanics Research Unit, GIGA in Silico Medicine, University of Liège, Belgium}

\begin{abstract}
A combination of reaction-diffusion models with moving-boundary problems yields a system in which the diffusion (spreading and penetration) and reaction (transformation) evolve the system's state and geometry over time. These systems can be used in a wide range of engineering applications. In this study, as an example of such a system, the degradation of metallic materials is investigated. A mathematical model is constructed of the diffusion-reaction processes and the movement of corrosion front of a magnesium block floating in a chemical solution. The corresponding parallelized computational model is implemented using the finite element method, and the weak and strong scaling behaviors of the model are evaluated to analyze the performance and efficiency of the employed high-performance computing techniques.
\end{abstract}


\end{frontmatter}



\section{Introduction}

Moving-boundary problems \cite{Crank1987} are a subset of the general concept of boundary-value problems which not only require the solution of the underlying partial differential equation (PDE), but also the determination of the boundary of the domain (or sub-domains) as part of the solution.  Moving-boundary problems are usually referred to as Stefan problems \cite{Crank1987} and can be used to model a plethora of phenomena ranging from phase separation and multiphase flows in materials engineering to bone development and tumor growth in biology. Reaction-diffusion systems are the  mathematical models in which the change of state variables occurs via transformation and spreading. These systems are described by a set of parabolic PDEs and can model a large number of different systems in science and engineering, for instance predator-prey models in biology and chemical components reactions in chemistry \cite{Grindrod1996}. Combining the reaction-diffusion systems with moving-boundary problems provides a way to study the systems in which the diffusion and reaction lead to the change of domain geometry. Such systems have great importance in various real-world scenarios in chemistry and chemical engineering as well as environmental and life sciences.

In this study, the material degradation phenomenon has been investigated as an example of a reaction-diffusion system with moving boundaries, in which the loss of material due to corrosion leads to movement of the interface of the bulk material and surrounding corrosion environment. More specifically, the degradation of magnesium (Mg) in simulated body fluid has been chosen as a case study. Magnesium has been chosen due to its growing usability as a degradable material in biomedicine, where it is usually used in  biodegradable implants for bone tissue engineering and cardiovascular applications \cite{Chen2014,Zhao2017}. The ultimate application of such a model can be then to study the degradation behavior of resorbable Mg-based biomaterials.

A wide range of different techniques has already been developed to study the moving interfaces in reaction-diffusion problems, which can be grouped into 3 main categories: 1) mesh elimination techniques, in which some elements are eliminated to simulate the interface movement (or loss of material in corrosion problems), 2) explicit surface representation, such as the arbitrary Lagrangian-Eulerian (ALE) method, which tracks the interface by moving a Lagrangian mesh inside an Eulerian grid, and 3) implicit surface tracking, in which an implicit criterion is responsible to define the moving interface during the reaction-diffusion process. Related to the aforementioned case study, studies performed by Gao et al. \cite{Gao2018} and Gastaldi et al. \cite{Gastaldi2011} are examples of the first group. Gao et al. \cite{Gao2018} have constructed a simulation of degradation using the mesh elimination technique. Gastaldi et al. \cite{Gastaldi2011} have developed a continuous damage (CD) model by using an explicit solver to study the degradation. The work of Grogan et al. \cite{Grogan2014} is an example of the second group as they have developed one of the first models to correlate the mass flux of the metallic ions in the biodegradation interface to the velocity of said interface. This was used to build an ALE model to explicitly track the boundary of the material during degradation. Studies of the third category are based more on mathematical modeling rather than available models in simulation software packages. This approach results in more flexibility and control over the implementation of the computational model. For instance, Wilder et al. \cite{Wilder2014} have derived a system of mathematical equations to study galvanic corrosion of metals, taking advantage of the level set method (LSM) to track the corrosion front. Bajger et al. \cite{Bajger2016} have used the definition of velocity of the biodegradation interface as the speed of the moving boundary in LSM, enabling them to track the geometrical changes of the material during degradation. A similar approach has been employed in this research.

Tracking the moving front at the diffusion interface requires high numerical accuracy of the diffusive state variables, which can be achieved using a refined computational grid. This makes the model computationally intensive, and as a consequence, implementing parallelization is an inevitable aspect of simulating such a model.  Such an approach enables the model to simulate large-scale systems with a large number of degrees of freedom (DOF) in 3D with higher performance and efficiency in high-performance computing (HPC) environments. In recent years, parallelization of diffusion-reaction systems simulation has been investigated, but the studies are mainly conducted for stochastic (statistical) models. For instance, Chen et al. \cite{Chen2017} have developed a parallel stochastic model for large-scale spatial reaction-diffusion simulation, and similarly, Arjunan et al. \cite{Arjunan2020} have developed a stochastic high-performance simulator for specific biological applications. Also as an example for massively parallel systems, Hallock et al. \cite{Hallock2014} have conducted a simulation of reaction-diffusion processes in biology using graphics processing units (GPUs). Although stochastic models have more parallel-friendly algorithms, explaining the underlying process, especially when it involves reaction-diffusion processes of chemistry and biology, is less complex and more universal using mechanistic (deterministic) models, which are based on well-developed mathematical models of continuous systems \cite{Kendall1999}. To the best of authors' knowledge, none of the previous contributions to the topic of reaction-diffusion systems with moving interfaces has employed parallelization techniques to increase the performance and speed of execution of the model without compromising the accuracy of the interface tracking. 

\begin{figure}
\center \includegraphics[width=12cm]{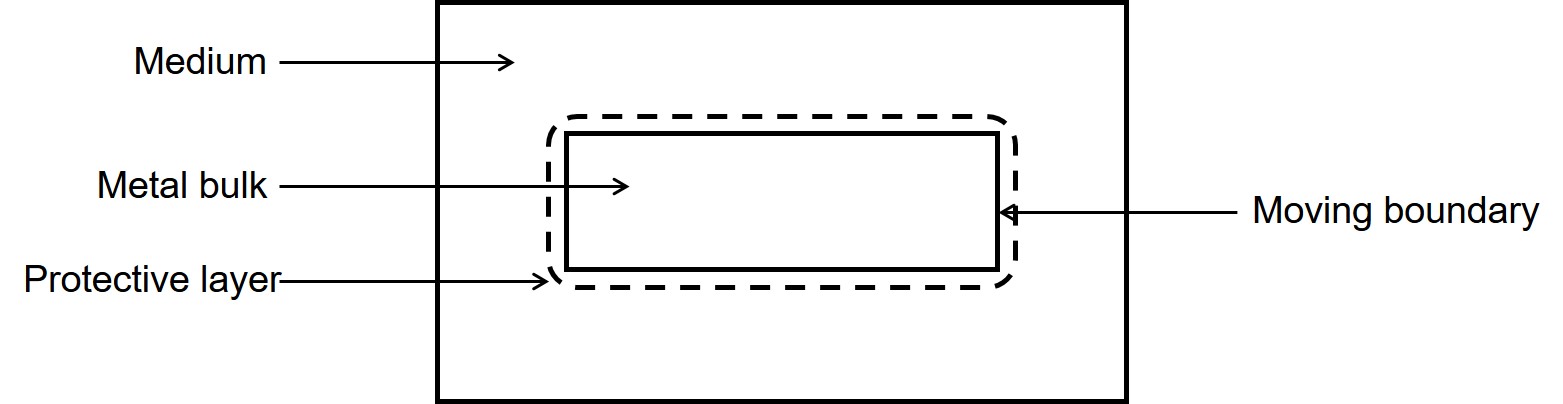}
\caption{A schematic representation of different components of the developed model for simulation of the degradation process with a moving front.} \label{fig:schematic}
\end{figure}

In the current study, we developed a mechanistic model of a reaction-diffusion system coupled with a moving interface problem. Improving the accuracy of the interface capturing requires a refined computational mesh, leading to a more computation-intensive simulation. To overcome this challenge and yield more interactable simulations, scalable  parallelization techniques were implemented making the model capable of being run on massively parallel systems to reduce the simulation time. The investigated case-study is the 
material degradation process. The developed model captures the release of metallic ions to the medium, formation of a protective film on the surface of the material, the effect of presented ions in the medium on the thickness of this protection layer, and tracking of the movement of the corrosion front (Fig. \ref{fig:schematic}). The interface tracking was performed using an implicit distance function that defined the position of the interface during degradation. This implicit function was obtained by constructing and solving a level set model. It is also worth noting that in a real-world application, such systems require a calibration (also called parameter estimation or inverse problem), in which the model should be simulated hundreds of times. This makes the parallelization even more crucial for these models.

\section{Background theory and model description}

Before elaborating the parallel implementation strategy, the mathematical model is briefly described in this section. The model is constructed based on the chemistry of degradation, starting from the previous work by Bajger et al. \cite{Bajger2016}, in which the ions can diffuse to the medium and react with each other.

\subsection{Chemistry of degradation}

In metals, degradation occurs through the corrosion process, which usually consists of electrochemical reactions, including anodic and cathodic reactions as well as the formation of side products \cite{Zheng2014}. 

For Mg, the corrosion reactions comprise the following steps \cite{Zheng2014}: first, the material is released as metallic ions and free electrons, which causes the volume of the bulk material to be reduced:
\begin{equation} \label{eq:oxidation_react}
\mathrm{Mg} \longrightarrow \mathrm{Mg}^{2+}+2 \mathrm{e}^{-}.
\end{equation}

The free electron reduces water to hydrogen gas and hydroxide ions:
\begin{equation} \label{eq:reduction_react}
2 \mathrm{H}_{2} \mathrm{O}+2 \mathrm{e}^{-} \longrightarrow \mathrm{H}_{2}+2 \mathrm{OH}^{-}.
\end{equation}

Then, with the combination of the metallic and hydroxide ions, a porous film is formed on the surface, slowing down the degradation rate by protecting the material underneath:
\begin{equation} \label{eq:formation_react}
\mathrm{Mg}^{2+}+2 \mathrm{OH}^{-} \longrightarrow \mathrm{Mg}(\mathrm{OH})_{2}.
\end{equation}

With the presence of some specific ions in the surrounding medium, such as chloride ions in a saline solution, the protective film might be broken partially, which contributes to an increase of the rate of degradation:
\begin{equation} \label{eq:break_react}
\mathrm{Mg}(\mathrm{OH})_{2}+2\mathrm{Cl}^{-} \longrightarrow \mathrm{Mg}^{2+} + 2\mathrm{Cl}^{-} + 2\mathrm{OH}^{-}.
\end{equation}

The degradation process of metals is a continuous repetition of the above reactions.

\subsection{Reaction-diffusion equation}

A reaction-diffusion partial differential equation can describe the state of a reaction-diffusion system by tracking the change of the concentration of the different components of the system over time \cite{Grindrod1996}. The equation is a parabolic PDE and can be expressed as
\begin{equation} \label{eq:react_diff}
\frac{\partial u}{\partial t}-\nabla \cdot [D \nabla u]=f(u)
\end{equation}
\noindent in which the change of the state variable $u=u(\mathbf{x},t), \mathbf{x} \in \Omega \subset \mathbb{R}^{3}$ is described as a combination of how it diffuses and how it is produced or eliminated via reactions. The term $f(u)$ is a smooth function that describes the reaction processes.

In the example used in this study, the state variable in Eq. \ref{eq:react_diff} is the concentration of effective chemical components involved in the degradation process, namely magnesium ions and the protective layer, denoted by $C_{\mathrm{Mg}}$ and $C_{\mathrm{Film}}$ respectively.
\begin{equation}
C_{\mathrm{Mg}} = C_{\mathrm{Mg}}(\mathbf{x},t), \quad C_{\mathrm{Film}} = C_{\mathrm{Film}}(\mathbf{x},t) \quad \mathbf{x} \in \Omega \subset \mathbb{R}^{3}
\end{equation}

\noindent  $\Omega$ is the whole domain of interest, including the bulk material and its surrounding medium. So, by assuming that the reaction rates of Eqs. \ref{eq:formation_react} and \ref{eq:break_react} are $k_1$ and $k_2$ respectively,  one can write the change of those state variables according to Eq. \ref{eq:formation_react} and Eq. \ref{eq:break_react} as
\begin{equation} \label{eq:pde_mg_primary}
\frac{\partial C_{\mathrm{Mg}}}{\partial t}=\nabla \cdot \left(D_{\mathrm{Mg}}^{e}   \nabla C_{\mathrm{Mg}} \right)-k_{1} C_{\mathrm{Mg}} +k_{2} C_{\mathrm{Film}} [\mathrm{Cl}]^{2}
\end{equation}

\begin{equation} \label{eq:pde_film_primary}
\frac{\partial C_\mathrm{Film}}{\partial t}=k_{1} C_{\mathrm{Mg}} -k_{2} C_{\mathrm{Film}} [\mathrm{Cl}]^{2}.
\end{equation}

\noindent We assumed that the concentration of the chloride ions is constant (denoted by $[\mathrm{Cl}]$ in the equation) and does not diffuse into the protective film. The missing part of the model described by Eqs. \ref{eq:pde_mg_primary} and \ref{eq:pde_film_primary} is the effect of the protective film on the reduction of the degradation rate. To this end, we defined a saturation term, $(1-\frac{C_{\mathrm{Film}}}{[\mathrm{Film}]_{\max }})$ for the concentration of Mg ions in the equations. By considering the film's porosity ($\epsilon$), the maximum concentration of the protective layer can be calculated based on its density ($\rho_{\mathrm{Mg}(\mathrm{OH})_{2}}$):
\begin{equation}
[\mathrm{Film}]_{\max }=\rho_{\mathrm{Mg}(\mathrm{OH})_{2}} \cdot (1-\epsilon).
\end{equation}

The defined saturation term acts as a function of space that varies between 0 and 1 in each point. By adding this term to the concentration of Mg ions, we can write
\begin{equation} \label{eq:pde_mg}
\frac{\partial C_{\mathrm{Mg}}}{\partial t}=\nabla \cdot \left(D_{\mathrm{Mg}}^{e}  \nabla C_{\mathrm{Mg}} \right)-k_{1} C_{\mathrm{Mg}}\left(1-\frac{C_{\mathrm{Film}}}{[\mathrm{Film}]_{\max }}\right) +k_{2} C_{\mathrm{Film}} [\mathrm{Cl}]^{2}
\end{equation}

\begin{equation} \label{eq:pde_film}
\frac{\partial C_\mathrm{Film}}{\partial t}=k_{1} C_{\mathrm{Mg}}\left(1-\frac{C_{\mathrm{Film}}}{[\mathrm{Film}]_{\max }}\right) -k_{2} C_{\mathrm{Film}} [\mathrm{Cl}]^{2}.
\end{equation}

Since the film is a porous layer and allows the ions to diffuse through it, the diffusion coefficient in Eq. \ref{eq:pde_mg} is a function of space and not a constant value (which is the reason for being denoted as $D_{\mathrm{Mg}}^{e}$). We can calculate this effective diffusion function by interpolating two values at any point: 1) $D_{\mathrm{Mg}}^{e} = D_{\mathrm{Mg}}$ when $C_\mathrm{Film} = 0$, and 2) $D_{\mathrm{Mg}}^{e} = \frac{\epsilon}{\tau} D_{\mathrm{Mg}}$ when $C_\mathrm{Film} = [\mathrm{Film}]_{\max }$, in which $\epsilon$ and $\tau$ are the porosity and tortuosity of the protective film, respectively. The interpolation leads to the effective diffusion function:

\begin{equation} \label{eq:diff_coeff}
D_{\mathrm{Mg}}^{e}=D_{\mathrm{Mg}}\left(\left(1-\frac{C_{\mathrm{Film}}}{[\mathrm{Film}]_{\max }}\right)+\frac{C_{\mathrm{Film}}}{[\mathrm{Film}]_{\max }} \frac{\epsilon}{\tau}\right).
\end{equation}

\subsection{Level set method}

\begin{figure}
\center \includegraphics[width=6cm]{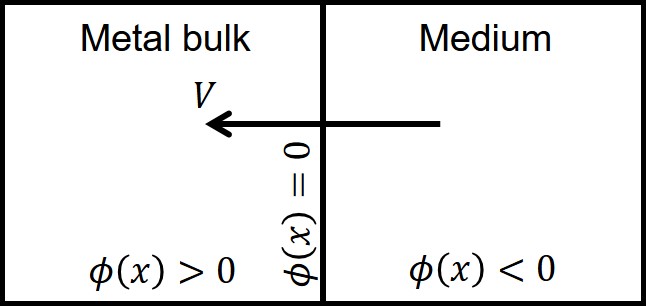}
\caption{A schematic representation of the implicit function definition in the current study. $V$ denotes the shrinkage speed of the interface due to degradation.} \label{fig:level_set_domain}
\end{figure}

The level set method is a methodology that allows moving interfaces to be described by an implicit function. In other words, the boundaries of domains can be tracked as a function instead of being explicitly defined. In the level set method, a signed distance function, $\phi = \phi(x,y,z,t)$, describes the distance of each point in space to the interface, and the zero isocontour of this function implies the interface \cite{RonaldFedkiw2002}. In the current study, this function was defined in a way that divides the domain into two subdomains: 1) the bulk material, in which the implicit function is positive ($\phi > 0$), and 2) the medium, in which the function is negative ($\phi < 0$). The interface is defined as the points in space where $\phi = 0$. Fig. \ref{fig:level_set_domain} shows a schematic representation of the solid-medium interface in the current study, in which the interface moves as the material degrades over time.

The level set equation defines this implicit function. The full level set equation can be written as \cite{RonaldFedkiw2002}:
\begin{equation} \label{eq:lsm_full}
\frac{\partial \phi}{\partial t}+{\overrightarrow{V^\mathrm{E}} \cdot \nabla \phi}+{\mathrm{V}^\mathrm{N}|\nabla \phi|}={b \kappa|\nabla \phi|}
\end{equation}
in which the terms correspond to temporal changes, external velocity field effect, normal direction motion, and curvature-dependent interface movement, respectively. 
$\overrightarrow{V^\mathrm{E}}$ is the external velocity field, and  $\mathrm{V}^\mathrm{N}$ is the magnitude of the interface velocity along the normal axis. In practical usage, some of the terms are  neglected. In this study, perfusion (rotation of the liquid around the bulk sample) is not considered, and the degradation rate does not depend on the curvature of the interface. As a result, by assuming that the interface moves in normal direction only, Eq. \ref{eq:lsm_full} can be simplified to
\begin{equation} \label{eq:lsm_simplified}
\frac{\partial \phi}{\partial t}+\mathrm{V}^\mathrm{N}|\nabla \phi|=0
\end{equation}
\noindent where $\mathrm{V}^\mathrm{N}$ is depicted in Fig. \ref{fig:level_set_domain}. The Rankine–Hugoniot equation can be used to calculate the interface velocity in mass transfer problems \cite{Scheiner2007}:

\begin{equation} \label{eq:rankine}
\left\{\mathbf{J}(x, t)-\left(c_{\mathrm{sol}}-c_{\mathrm{sat}}\right) \mathrm{V}(x, t)\right\} \cdot n=0
\end{equation}
\noindent in which $\mathbf{J}$ is the mass flux, $c_{\mathrm{sol}}$ is the concentration of the material in the bulk part (i.e. its density), and $c_{\mathrm{sat}}$ is the concentration at which the material (here, the ions) saturates through the medium. So, for the investigated Mg degradation problem, Eq. \ref{eq:rankine} will be:
\begin{equation} \label{eq:rankine_mg}
D_{\mathrm{Mg}}^{e} \nabla_{n} C_\mathrm{Mg}-\left([\mathrm{Mg}]_{\mathrm{sol}}-[\mathrm{Mg}]_{\mathrm{sat}}\right) \mathrm{V}^\mathrm{N}=0.
\end{equation}

Inserting the obtained velocity of Eq. \ref{eq:rankine_mg} into Eq. \ref{eq:lsm_simplified} yields
\begin{equation} \label{eq:lsm_final}
\frac{\partial \phi}{\partial t}-\frac{D_{\mathrm{Mg}}^{e} \nabla_{n} C_\mathrm{Mg}}{[\mathrm{Mg}]_{\mathrm{sol}}-[\mathrm{Mg}]_{\mathrm{sat}}}|\nabla \phi|=0.
\end{equation}

Eq. \ref{eq:lsm_final} is the final formulation of the level set equation in the current study, which alongside Eqs. \ref{eq:pde_mg} and \ref{eq:pde_film} forms the mathematical model of degradation of Mg with a moving interface. 

\section{Methodology of model implementation}

The developed mathematical model comprised of Eqs. \ref{eq:pde_mg}, \ref{eq:pde_film}, and \ref{eq:lsm_final} cannot be solved using analytical techniques. The alternative approach in these scenarios is solving the derived PDEs numerically. In this study, we used a combination of finite element and finite difference methods to solve the aforementioned equations. In the following section, only the process to obtain the solution of Eq. \ref{eq:pde_mg} is described in detail, but the other PDEs were solved using the same principle. Although the adopted finite element method is standard, we elaborate on its derivation to clarify the bottlenecks of the later-discussed implementation.

\subsection{Finite element discretization (bottleneck of the algorithm)}

In order to solve Eq. \ref{eq:pde_mg} numerically, we used a finite difference scheme for the temporal term and a finite element formulation for the spatial terms. For simplicity of writing, notations of variables are changed, so $C_\mathrm{Mg}$ is represented as $u$ (the main unknown state variable to find), $C_\mathrm{Film}$ is denoted by $p$, $[\mathrm{Cl}]$ is denoted by $q$, and the saturation term $(1-\frac{F}{F_{\max }})$ is denoted by $s$. By doing this, Eq. \ref{eq:pde_mg} can be written as
\begin{equation} \label{eq:pde_changed_not}
\frac{\partial u}{\partial t}=\nabla \cdot (D   \nabla u)-k_{1} s u+k_{2} p q^{2}.
\end{equation}

To obtain the finite element formulation, the weak form of derived PDE is required. In order to get this, we define a space of test functions and then, multiply each term of the PDE by any arbitrary function as a member of this space. The test function space is 
\begin{equation} \label{eq:function_space}
\mathcal{V}=\left\{v(\mathbf{x}) | \mathbf{x} \in {\Omega}, v(\mathbf{x}) \in \mathcal{H}^{1}(\Omega), \text { and } v(\mathbf{x})=0 \text { on } \Gamma\right\}
\end{equation}
in which the $\Omega$ is the domain of interest, $\Gamma$ is the boundary of $\Omega$, and $\mathcal{H}^{1}$ denotes the Sobolev space of the domain $\Omega$, which is a space of functions whose derivatives are square-integrable functions in $\Omega$. The solution of the PDE belongs to a trial function space, which is similarly defined as 
\begin{equation} \label{eq:trial_domain}
\mathcal{S}_{t}=\left\{u(\mathbf{x}, t) | \mathbf{x} \in \Omega, t>0, u(\mathbf{x}, t) \in \mathcal{H}^{1}(\Omega), \text { and } \frac{\partial u}{\partial n}=0 \text { on } \Gamma\right\}.
\end{equation}

Then, we multiply Eq. \ref{eq:pde_changed_not} to an arbitrary function $v \in \mathcal{V}$:

\begin{equation}
\frac{\partial u}{\partial t} v=\nabla \cdot (D  \nabla u) v-k_{1} s u v+k_{2} p q^{2} v.
\end{equation}

\noindent Integrating over the whole domain yields:
\begin{equation} \label{eq:int_first}
\int_{\Omega} \frac{\partial u}{\partial t} v d \omega=\int_{\Omega} \nabla \cdot (D  \nabla u) v d \omega-\int_{\Omega} k_{1} s u v d \omega+\int_{\Omega} k_{2} p q^{2} v d \omega.
\end{equation}

\noindent The diffusion term can be split using the integration by parts technique:
\begin{equation} \label{eq:int_part}
\int_{\Omega} \nabla \cdot (D  \nabla u) v d \omega = \int_{\Omega} \nabla \cdot[v(D  \nabla u)] d \omega-\int_{\Omega} (\nabla v) \cdot(D  \nabla u) d \omega 
\end{equation}

\noindent in which the second term can be converted to a surface integral on the domain boundary by applying the Green's divergence theory:
\begin{equation} \label{eq:divergence}
\int_{\Omega} \nabla \cdot[v(D  \nabla u)] d \omega = \int_{\Gamma} D v \frac{\partial u}{\partial n} d \gamma.
\end{equation}

\noindent For the temporal term, we use the finite difference method and apply a first-order backward Euler scheme for discretization, which makes it possible to solve the PDE implicitly: 
\begin{equation} \label{eq:backward}
\frac{\partial u}{\partial t} = \frac{u-u^{n}}{\Delta t}
\end{equation}
\noindent where $u^n$ denotes the value of the state variable in the previous time step (or initial condition for the first time step). Inserting Eqs. \ref{eq:int_part}, \ref{eq:divergence}, and \ref{eq:backward} into Eq. \ref{eq:int_first} yields:
\begin{equation}
\int_{\Omega} \frac{u-u^{n}}{\Delta t} v d \omega=\int_{\Gamma} D v  \frac{\partial u}{\partial n} d \gamma-\int_{\Omega} D  \nabla u \cdot \nabla v d \omega-\int_{\Omega} k_{1} s u v d \omega+\int_{\Omega} k_{2} p q^{2} v d \omega.
\end{equation}

\noindent The surface integral is zero because there is a no-flux boundary condition on the boundary of the computational domain (defined in the trial function space according to Eq. \ref{eq:trial_domain}). By reordering the equation, we get the weak form of Eq. \ref{eq:pde_changed_not}:
\begin{equation}  \label{eq:weak_general}
\int_{\Omega} {u} v d \omega+\int_{\Omega} \Delta t D  \nabla u \cdot  \nabla v d \omega+\int_{\Omega} \Delta tk_{1} s u v d \omega=\int_{\Omega} {u^{n}} v d \omega+\int_{\Omega} \Delta t k_{2} p q^{2} v d \omega.
\end{equation}



So, the problem is finding a function $u(t) \in \mathcal{S}_{t}$ such that for all $v \in \mathcal{V}$ Eq. \ref{eq:weak_general} would be satisfied. By defining a linear functional $(f, v) =\int_{\Omega} f v d \omega$ and encapsulating the independent concentration terms into $f^{n} = pq^2$, Eq. \ref{eq:weak_general} can be simplified as:
\begin{equation}
(u, v)[1+\Delta t k_1 s]+\Delta t(D \nabla u, \nabla v)=\left(u^{n}, v\right)+\Delta t\left(f^{n}, v\right)
\end{equation}


\noindent which can be further converted to the common form of the weak formulation of time-dependent reaction-diffusion PDEs by multiplying to a new coefficient $\alpha = \frac{1}{1+\Delta t k_1 s}$: 

\begin{equation} \label{eq:weak_compact}
(u, v)+ \alpha \Delta t(D \nabla u, \nabla v)=\alpha \left(u^{n}, v\right)+ \alpha \Delta t\left(f^{n}, v\right).
\end{equation}

One can approximate the unknown function $u$ in Eq. \ref{eq:weak_compact} by $u(x) \approx \sum_{i=0}^{N} c_{i} \psi_{i}(x)$, where the $\psi_{i}$ are the basis functions used to discretize the function space, and $c_0,\ldots,c_N$ are the unknown coefficients. The finite element method uses Lagrange polynomials as the basis function and discretizes the computational domain using a new function space $\mathcal{V}_h$ spanned by the basis functions $\left\{\psi_{i}\right\}_{i \in \mathcal{I}_{s}}$, in which $\mathcal{I}_{s}$ is defined as $\mathcal{I}_{s}=\{0, \ldots, N\}$, where $N$ denotes the degrees of freedom in the computational mesh. The computational mesh discretizes the space into a finite number of elements, in each of which the $\psi_{i}$ is non-zero inside the $i$th element and zero everywhere else. In this study, 1st order Lagrange polynomials were used as the basis functions to define the finite element space.

%

In order to derive a linear system of equations for obtaining the unknown coefficients $c_j$, we define 
\begin{equation} \label{eq:u_definition}
u=\sum_{j=0}^{N} c_{j} \psi_{j}(\boldsymbol{x}), \quad u^{n}=\sum_{j=0}^{N} c_{j}^{n} \psi_{j}(\boldsymbol{x})
\end{equation}

\noindent as the definition of the unknown function $u$ and its value in the previous time step $u^n$. We then insert it into Eq. \ref{eq:weak_compact}, which yields the following equation for each degree of freedom $i=0, \ldots, N$, where the test functions are selected as $v = \psi_i$:
\begin{equation} \label{eq:weak_sum_all}
\sum_{j=0}^{N}\left(\psi_{i}, \psi_{j}\right) c_{j} + \alpha \Delta t \sum_{j=0}^{N}\left(\nabla \psi_{i}, D \nabla \psi_{j}\right) c_{j} =\sum_{j=0}^{N} \alpha \left(\psi_{i}, \psi_{j}\right) c_{j}^{n}+\alpha\Delta t\left(f^{n}, \psi_{i}\right).
\end{equation}

Eq. \ref{eq:weak_sum_all} is a linear system
\begin{equation} \label{eq:linear_system}
\sum_{j} A_{i, j} c_{j}=b_{i}
\end{equation}
with
\begin{equation} \label{eq:system_a}
A_{i, j}=\left(\psi_{i}, \psi_{j}\right) + \alpha \Delta t \left(\nabla \psi_{i}, D \nabla \psi_{j}\right)
\end{equation}
\begin{equation} \label{eq:system_b}
b_{i}=\sum_{j=0}^{N}\alpha \left(\psi_{i}, \psi_{j}\right) c_{j}^{n}+\alpha \Delta t\left(f^{n}, \psi_{i}\right)
\end{equation}

%

By solving Eq. \ref{eq:linear_system} and substituting the obtained $c$ in Eq. \ref{eq:u_definition}, $u$ ($C_{\mathrm{Mg}}$ in the example in this study) can be calculated in the current time step. As stated before, the same approach can be applied to Eq. \ref{eq:pde_film} and Eq. \ref{eq:lsm_final} to get $C_{\mathrm{Film}}$ and $\phi$. This procedure is repeated in each time step to compute the values of $C_{\mathrm{Mg}}$, $C_{\mathrm{Film}}$, and $\phi$ over time. 

A common practice to save time for solving Eq. \ref{eq:linear_system}  for a constant time step size is to compute the left-hand side matrix ($A$ in Eq. \ref{eq:system_a}) once and compute only the right-hand side vector of the equation at each time iteration. But in this case, although the time step size is fixed, due to the presence of the $\alpha$ coefficient, the matrix changes along the time. The $\alpha$ coefficient is not constant and should be updated in each time step because it depends on the penalization term $s$ (which is a function of the concentration of the film as can be seen by comparing Eq. \ref{eq:pde_mg} and Eq. \ref{eq:pde_changed_not}). In addition to this, the diffusion coefficient is not constant (Eq. \ref{eq:diff_coeff}), making the second term in Eq. \ref{eq:system_a} non-constant even in the absence of $\alpha$ coefficient. Consequently, the left-hand side matrix of the Eq. \ref{eq:linear_system} cannot be computed before the start of the main time loop, and computing it in each time step is an extra but inevitable computational task in comparison to similar efficient and high-performance finite element implementations. This contributes to a slower algorithm for solving the aforementioned PDEs.

\subsection{Implementation and parallelization}

The model was implemented in FreeFEM \cite{Hecht2012}, which is an open-source PDE solver to facilitate converting the weak formulation (Eq. \ref{eq:weak_general}) to a linear system $Ax=b$ (with $A$ from Eq. \ref{eq:system_a} and $b$ from Eq. \ref{eq:system_b}). The computational mesh was generated using Netgen \cite{Schoeberl1997} in the SALOME platform \cite{Ribes2007} by a set of linear tetrahedral elements, and all the other preprocessing steps were performed in FreeFEM. The mesh was adaptively refined on the material-medium interface in order to increase the accuracy of the level set model. Computing the diffusion solely in the medium domain causes oscillations close to the interface, and to prevent this, the mass lumping feature of FreeFEM was employed. Postprocessing of the results was carried out using  Paraview \cite{Ahrens2005}.


The main parallelization approach for the current study was domain decomposition, in which the mesh is split into smaller domains (can be overlapping or non-overlapping), and the global solution of the linear system is achieved by solving the problem on each smaller local mesh. What really matters in this approach is providing virtual boundary conditions to the smaller sub-domains by ghost elements, transferring neighboring sub-domain solutions \cite{Badri2018}. As a result, a high-performance parallelism is feasible by assigning each sub-domain to one processing unit. 

In computational science, preconditioning is widely used to enhance the convergence, which means instead of directly working with a linear system $Ax=b$, one can consider the preconditioned system \cite{Daas2019AMS}:
\begin{equation} \label{eq:precond_system}
M^{-1} A x=M^{-1} b
\end{equation}
in which the $M^{-1}$ is the preconditioner. In the current study, we considered this approach for both the domain composition and the solution of the linear system. We opted to use an overlapping Schwarz method for domain decomposition, in which the mesh is first divided into a graph of $N$ non-overlapping meshes using METIS (or ParMETIS) \cite{METIS1998}. Then, by defining a positive number $\delta$, the overlapping decomposition $\left\{\mathcal{T}_{i}^{\delta}\right\}_{1 \leqslant i \leqslant N}$ can be created recursively for each sub-mesh $\left\{\mathcal{T}_{i}\right\}_{1 \leqslant i \leqslant N}$ by adding all adjacent elements of $\mathcal{T}_{i}^{\delta-1}$ to it. Then, the finite element space $\mathcal{V}_{h}$ (Eq. \ref{eq:function_space}) can be mapped to the local space $\left\{\mathcal{V}_{i}^{\delta}\right\}_{1 \leqslant i \leqslant N}$ by considering the restrictions $\left\{R_{i}\right\}_{1 \leqslant i \leqslant N}$ and a local partition of unity $\left\{D_{i}\right\}_{1 \leqslant i \leqslant N}$ such that:
\begin{equation} \label{eq:restrict}
\sum_{j=1}^{N} R_{j}^{\top} D_{j} R_{j}=I_{n \times n}
\end{equation}
where $I$ and $n$ denote identity matrix and the global number of unknowns, respectively \cite{Dolean2015}.

\begin{figure}[ht]
\center \includegraphics[width=5cm]{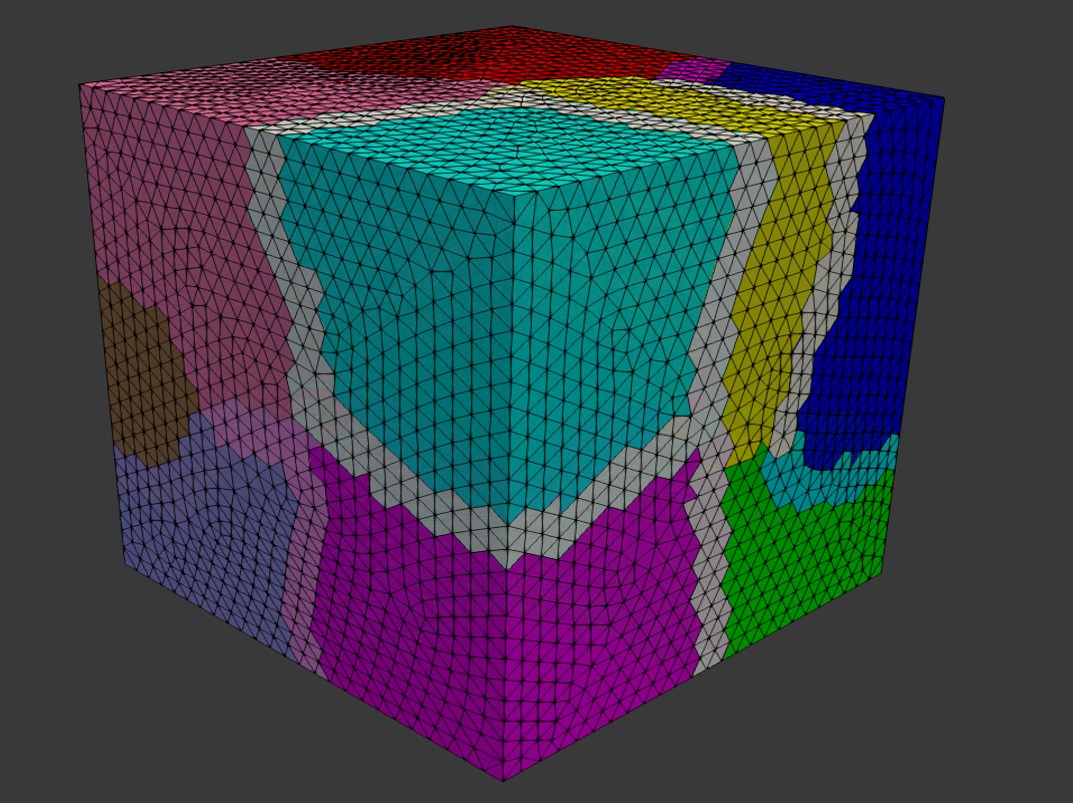}
\caption{Overlapping domain decomposition in the current study. Each color shows a separate sub-domain, and the narrow lighter bands are the overlapped regions.} \label{fig:overlap}
\end{figure}

\begin{figure}[ht]
\center \includegraphics[width=11cm]{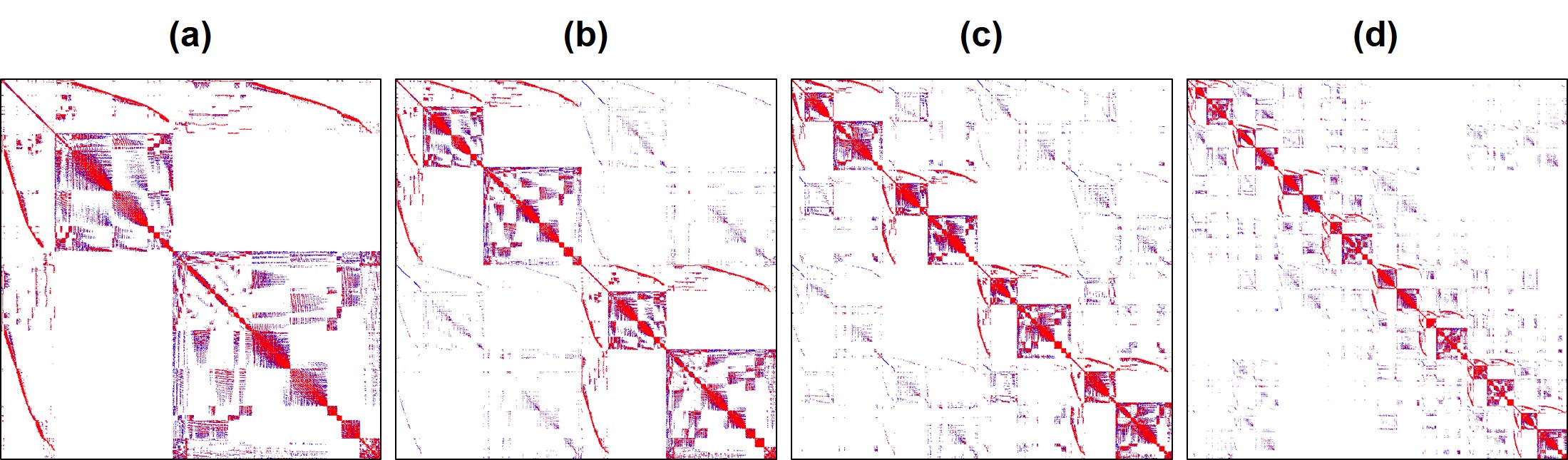}
\caption{Comparison of the sparsity patterns (highlighting non-zero elements) of the global matrix A for a different number of decomposed domains a: 1 domain b: 2 sub-domains c: 4 sub-domains d: 8 sub-domains.} \label{fig:sparsity_pattern}
\end{figure}

In this study, we decomposed the mesh by using the one-level preconditioner Restricted Additive Schwarz (RAS):
\begin{equation} \label{eq:ras}
M_{\mathrm{RAS}}^{-1}=\sum_{i=1}^{N} R_{i}^{\top} D_{i} A_{i}^{-1} R_{i}
\end{equation}
in which $\left\{A_{i}\right\}_{1 \leqslant i \leqslant N}$ is the local operator of the sub-matrices \cite{Dolean2015}. For this purpose, we took advantage of the HPDDM package interface in FreeFEM \cite{Jolivet2013}. The partitioned mesh is shown in Fig. \ref{fig:overlap}. The effect of the construction of these local sub-domains on the sparsity pattern of the global matrix is also depicted in Fig. \ref{fig:sparsity_pattern}. The global matrix is a sparse matrix according to Eq. \ref{eq:system_a} and the definition of the basis function $\psi$.

Generally, two categories of methods have been used to solve a large linear system of equations on parallel machines: direct solvers (e.g. MUMPS \cite{MUMPS1}) and iterative solvers (e.g. GMRES \cite{Saad1986}). While direct solvers are quite robust, they suffer from the memory requirement problem on large systems. Inversely, iterative solvers are quite efficient on memory consumption, but similar to other iterative approaches, they are not very reliable in some cases \cite{Saad2003}. Direct solvers modify the matrix by factorization (e.g. Cholesky decomposition), but an iterative solver does not manipulate the matrix and works solely using basic algebraic operations. However, for an efficient usage of iterative solvers, a proper preconditioner is crucial \cite{Saad2003}. By evaluating and comparing the performance of the aforementioned methods for the current model, we decided to use an iterative approach using the Krylov subspaces (KSP) method, in which we preconditioned the equation using a proper preconditioner (Eq. \ref{eq:precond_system}) and then solved it with an iterative solver.

Krylov methods have been frequently used by researchers as robust iterative approaches to parallelism \cite{Ipsen1998}. What matters in this regard is ensuring proper scaling of the parallelized algorithm for both the assembling of the matrices and the solution of the linear system of equations. One good solution to this challenge is taking advantage of HPC-ready mathematical libraries to achieve efficient distributed-memory parallelism through the Message Passing Interface (MPI). In the current study, we used the PETSc library \cite{petsc}, which provides a collection of high-performance preconditioners and solvers for this purpose. 

In order to yield the highest performance, a variety of different combinations of KSP types and preconditioners were evaluated, such as Conjugate Gradients (CG) \cite{hestenes1952}, Successive Over-Relaxation (SOR) \cite{habetler1961}, block Jacobi, and Algebraic Multigrid (AMG) \cite{mccormick1987}, to name a few. The best performance for the reaction-diffusion system model was achieved using the HYPRE preconditioner \cite{Falgout2002} and the GMRES solver \cite{Saad1986}. This was the combination used for all the performance analysis tests.

\subsection{Level set issues}

As mentioned before, in order to track the interface of the bulk material and the surrounding fluid, an implicit signed distance function is defined as the solution of Eq. \ref{eq:lsm_final}. This equation can be solved using the aforementioned finite element discretization, but in a practical implementation, there are usually a couple of problems associated with this PDE.

The first issue is defining $D_{\mathrm{Mg}}^{e}$ and $\nabla_{n} C_\mathrm{Mg}$ on the moving interface. To ensure correct boundary conditions for Eq. \ref{eq:rankine_mg}, the value of $C_\mathrm{Mg}$ is set constant on the whole bulk material by using the penalty method. As a result, the implicit interface is not necessarily aligned on the computational mesh. Although this is a beneficial fact for the interface tracking, it inserts the problem of overestimation of $C_\mathrm{Mg}$ on the nodes close to the interface, which makes it difficult to calculate $\nabla_{n} C_\mathrm{Mg}$ on these nodes correctly. The same problem exists for calculating $D_{\mathrm{Mg}}^{e}$. To overcome this issue, the values of $C_\mathrm{Mg}$ and $D_{\mathrm{Mg}}^{e}$ are calculated at the distance $h$ from the interface in the normal direction (towards the medium), where $h$ is the edge size of the smallest element of the computational mesh.

The next issue is a well-known problem of the level set method: if the velocity of the interface is not constant (as in Eq. \ref{eq:lsm_full}), the level set function $\phi$ may become distorted by having too flat or too steep gradients close to the moving front. This could cause unwanted movements of the interface. The problem becomes even worse when the distance function is advected. A solution to this issue is re-initializing the distance function in each time step (re-distancing), but this operation requires solving a new PDE. From numerical investigations, it has been observed that this operation inserts new errors in the numerical computation of the level set equation \cite{Russo2000}. This can be resolved by improving the method of reconstruction of the distance function \cite{Russo2000}. 

However, re-initialization results in another issue on a massively parallel implementation: as the mesh is partitioned into smaller sub-meshes, it is not feasible anymore to evaluate the distance to the interface globally on each sub-domain. As a result, the inverse process of domain decomposition should be taken to assemble the mesh again. This can be done by the restriction matrix and the partition of unity (defined in Eqs. \ref{eq:restrict} and \ref{eq:ras}), but it is rather a very inefficient procedure regarding the parallelization of the simulation and results in a long execution time in each time step.

In the current study, the distance function was not advected, and no distortion was observed either, so the distance function $\phi$ was initialized only once at the beginning of the simulation. This also removed the need for inverting the decomposition process.

\subsection{Simulation setup}

In order to verify the performance of the developed model, a degradation experiment was reconstructed in-silico, in which the degradation of a block of Mg (with the size of $13mm \times 13mm \times 4 mm$) was investigated in a simulated body fluid solution. All the experimental parameter data (used to setup the simulation), as well as the degradation rates (used to calibrate and validate the numerical model) were extracted from Mei et al.  \cite{Mei2019}. 

As can be seen in Eqs. \ref{eq:oxidation_react} and \ref{eq:reduction_react}, each mole removed from the Mg block corresponds to one mole of the produced hydrogen. As a result, instead of a direct measurement of mass loss, one can collect and measure the amount of produced hydrogen to monitor the degradation rate. This is a common way of reporting degradation in this type of studies \cite{Abidin2013}. In order to get this quantity out of the developed model, we used the level set model output. The total mass loss of Mg at each desired time can be calculated based on the movement of the corrosion front:

\begin{equation} \label{eq:mass_loss}
\mathrm{Mg}_{\mathrm{lost}}=\int_{\Omega_{+}(t)} \mathrm{Mg}_{\mathrm{solid}} \mathrm{d} V-\int_{\Omega_{+}(0)} \mathrm{Mg}_{\mathrm{solid}} \mathrm{d} V_{0}
\end{equation}
where $\Omega_{+}(t)=\{\mathbf{x}: \phi(\mathbf{x}, t) \geq 0\}$. It is worth noting that this integration should be performed by ignoring the ghost elements generated in the mesh partitioning process, otherwise the calculated material loss will be higher than the real value. Then, the amount of formed hydrogen gas can be calculated based on the ideal gas law:

\begin{equation} \label{eq:evolv_hydr}
H_{f}=\frac{\mathrm{Mg}_{\mathrm{lost}}}{\mathrm{Mg}_{\mathrm{mol}}} \frac{R T}{P A}
\end{equation}
in which $R$ is the universal gas constant, $P$ is the pressure, $T$ is the solution temperature, $A$ is the exposed corrosion surface area (which can be computed using the level set function), and $\mathrm{Mg}_{\mathrm{mol}}$ is the molar mass of Mg. Plotting a comparison of the predicted values of predicted and experimentally obtained values of hydrogen can show the overall validity of the  mathematical model because both the diffusion-reaction equations and the level set equation contribute to the prediction made by the computational model.

\begin{figure}
\center \includegraphics[width=13cm]{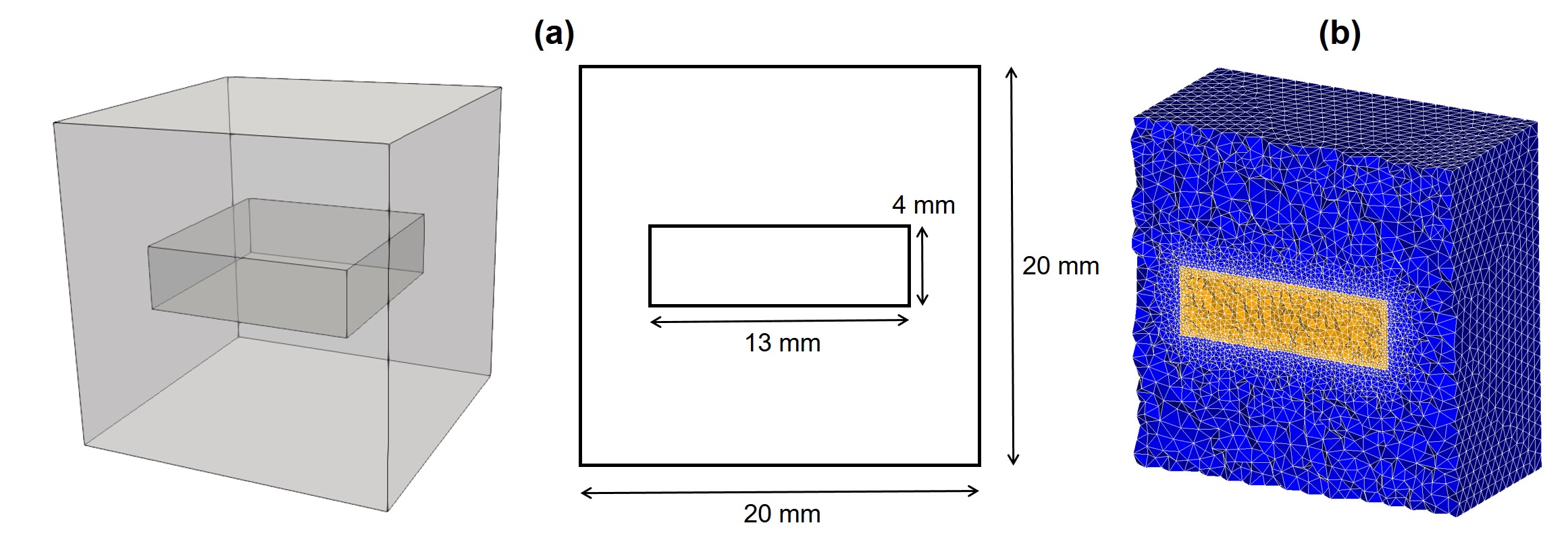}
\caption{Representation of the experimental set-up simulated  to perform numerical validation of the developed model and evaluate parallel performance. a) A cuboid of Mg (with the size of $13mm \times 13mm \times 4 mm$) is floating inside a simulated body fluid solution to investigate the degradation process, b) a cross-section of the computational mesh, refined on the metal-medium interface to increase the interface capturing accuracy.} \label{fig:simulation_setup}
\end{figure}

The geometry of the simulation experiment is depicted in Fig. \ref{fig:simulation_setup}. Based on this geometry, an Eulerian computational mesh was constructed by generating tetrahedral elements on the whole domain, including the Mg block and the medium. This resulted in $830\,808$ elements with a total of $143\,719$ DOFs for each PDE. Model parameters and material properties were obtained from Bajger et al. \cite{Bajger2016}. The diffusion coefficient of Mg was calculated using an inverse problem setup in which a Bayesian optimization process \cite{Mockus1989} was used to run the simulation code multiple times and minimize the difference of the model output and the experimental data reported by Mei. et al. \cite{Mei2019}. After a time step sensitivity/convergence study, the time step value was set to 0.025 hours.

\subsection{Performance analysis}

To investigate the performance and scaling behavior of the implemented parallel code, we conducted a set of weak-scaling and strong-scaling tests on the computational model. To do this, the time required to solve each PDE in each time step was measured in a simulation. This acted as a rough estimation of the time required in each time step because it ignores all the other factors contributing to speedup results such as communication costs, load imbalance, limited memory bandwidth, and parallelization-caused overhead.

\begin{figure}[ht]
\center \includegraphics[width=13cm]{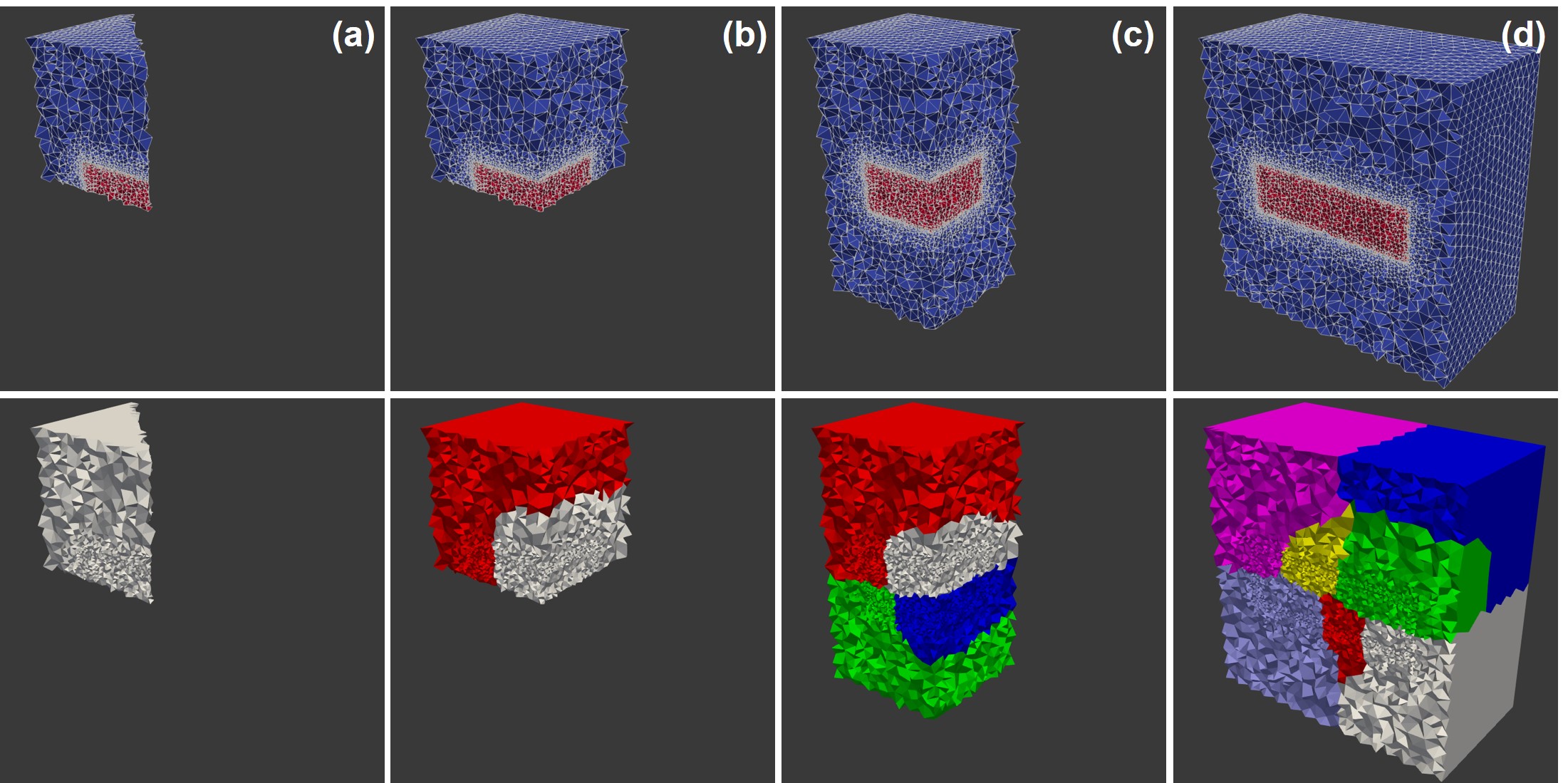}
\caption{Models used for weak-scaling, in which the number of elements was doubled each time while doubling the number of computational cores. Upper row: actual computational domain in which colors show the medium (blue) and the material block (red). Lower row: domain decomposition for parallelization, colors show different decomposed mesh parts (distributed to different MPI processing units). Each column corresponds to a different simulation with a: 1 MPI unit, b: 2 MPI units, c: 4 MPI units, and d: 8 MPI units.} \label{fig:weak_scaling_models}
\end{figure}

Weak-scaling was evaluated by dividing the computational domain into smaller sub-domains (each of which was $\frac{1}{16}$ of the whole domain, Fig. \ref{fig:weak_scaling_models}) and conducting simulation experiments with 1, 2, 4, and 8 computational cores in  a way  that the number of processors corresponded to the number of employed sub-domains. In Fig. \ref{fig:weak_scaling_models} the upper row shows different domains as an accumulation of the smaller divisions, and the lower row shows the corresponding domain decomposition for parallel computing by depicting each processing unit in a different color. In fact, it demonstrates the concept of increasing the number of MPI processing units as we increase the size of the problem.

After calculating the speedup of each test (by comparing the differences in execution time), we can use Gustafson’s law \cite{Gustafson1988} to calculate the sequential and parallelizable portion of computation in the current implementation in weak-scaling evaluation:
\begin{equation} \label{eq:weak_scaling}
\mathrm{Speedup} = f + (1-f) \times N
\end{equation}
where $N$ is the total number of computational cores, $f$ is the fraction of operations in the computation that are sequential, and as a result, $1-f$ is the fraction of the execution time spent on the parallelizable part.

The strong-scaling evaluation was performed using the entire domain. The evaluation was done using 1, 8, 10, 16, 40, 60, and 90 MPI cores. In strong-scaling, Amdahl’s law \cite{Amdahl1967} is used to calculate the portion of the algorithm that runs in parallel:
\begin{equation} \label{eq:strong_scaling}
\mathrm{Speedup} = \frac{1}{f + \frac{1-f}{N}}
\end{equation}
in which the parameters are the same as Eq. \ref{eq:weak_scaling}.

\subsection{Compute environment}

Simulations were conducted on the VSC (Flemish Supercomputer Center) supercomputer with the availability of Intel CPUs in three different micro-architectures: Ivy Bridge, Haswell, and Skylake. Due to a better performance, the strong and weak-scaling measurements were solely performed on the Skylake nodes. On this supercomputer, we made use of 3 nodes, 36 cores each, with 576 GB of the total memory, each node holding 2 Intel Xeon Gold 6132 CPUs with a base clock speed of 2.6 GHz. The supercomputer uses CentOS 7.6.1810 with kernel version 3.10.0. For interprocess communication, Intel's MPI implementation 2018 was used.

\section{Results}

\subsection{Numerical simulation results}

The performed numerical simulation produces the output of three main quantities: the concentration of the Mg ions in the medium (as the solution of Eq. \ref{eq:pde_mg}), the concentration of the protective film (as the solution of Eq. \ref{eq:pde_film}), and the level set function values at each element (as the solution of Eq. \ref{eq:lsm_final}). In addition to this, a quantitative prediction of the mass loss is also generated according to Eqs. \ref{eq:mass_loss} and \ref{eq:evolv_hydr}.

\begin{figure}
\center \includegraphics[width=12cm]{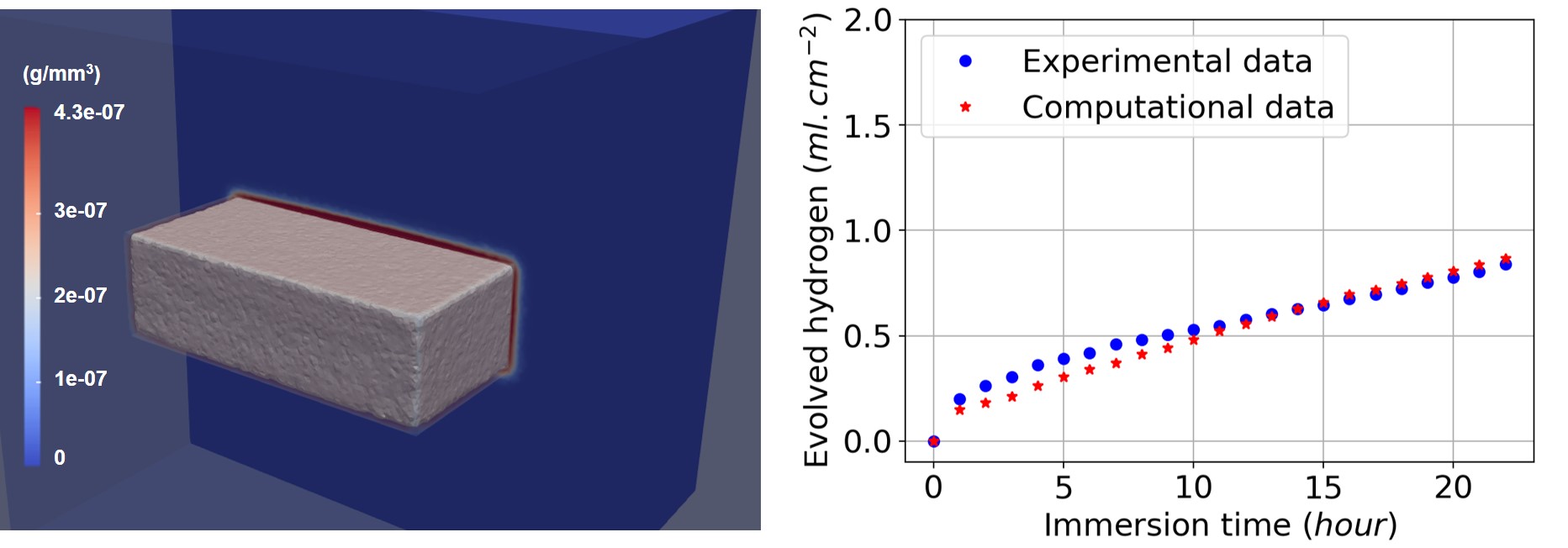}
\caption{Numerical simulation result. Left: formation of a protective layer on the surface of the Mg block (red region). Right: comparison of the produced hydrogen (a surrogate for the material loss) in the computational model and the experimental data, which is a validation of the full model as both the reaction-diffusion equations and the level-set equation are involved in the computation of this quantity.} \label{fig:numerical_results}
\end{figure}

  In order to have quantitative predictions, the coefficients of Eqs. \ref{eq:pde_mg} and \ref{eq:pde_film} (diffusion rates and reaction rates) should be calibrated using an inverse problem. Fig. \ref{fig:numerical_results} shows the results produced by the computational model after this parameter estimation stage. A narrow layer of the protective film is formed on the surface of the Mg block, and the volume of produced hydrogen gas is compared with values obtained from experiments. Additionally, by plotting the zero iso-contour of the level set function, we can obtain the shape of the material block as it degrades during the degradation process (i.e. tracking the moving corrosion front). This is depicted by the grey surface in Fig. \ref{fig:numerical_results}.  

\subsection{Weak and strong scaling results}

\begin{figure}
\center \includegraphics[width=14cm]{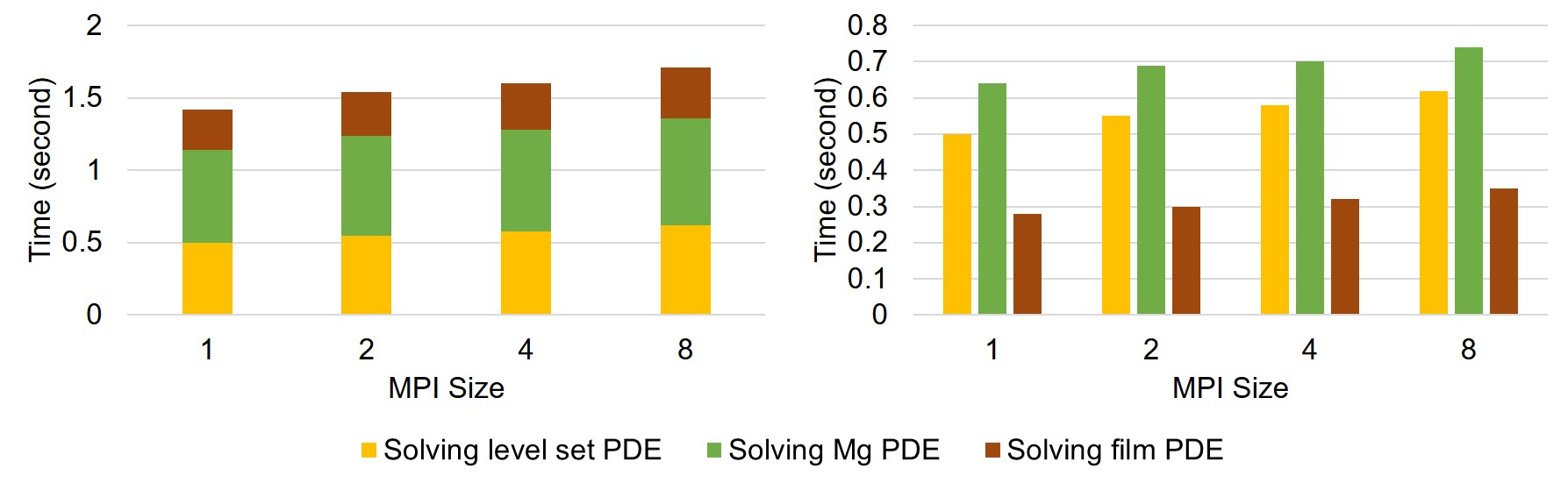}
\caption{Weak-scaling test result. Results are broken down into contributions for each PDE, which are plotted  cumulatively and separately in the left and right plot, respectively.} \label{fig:weak_scaling_results}
\end{figure}

Weak-scaling results are plotted in Fig. \ref{fig:weak_scaling_results}, in which the execution time of each time step is broken down into the time spent on each PDE. The results show good scalability of the parallel implementation. 

\begin{figure}[ht]
\center \includegraphics[width=14cm]{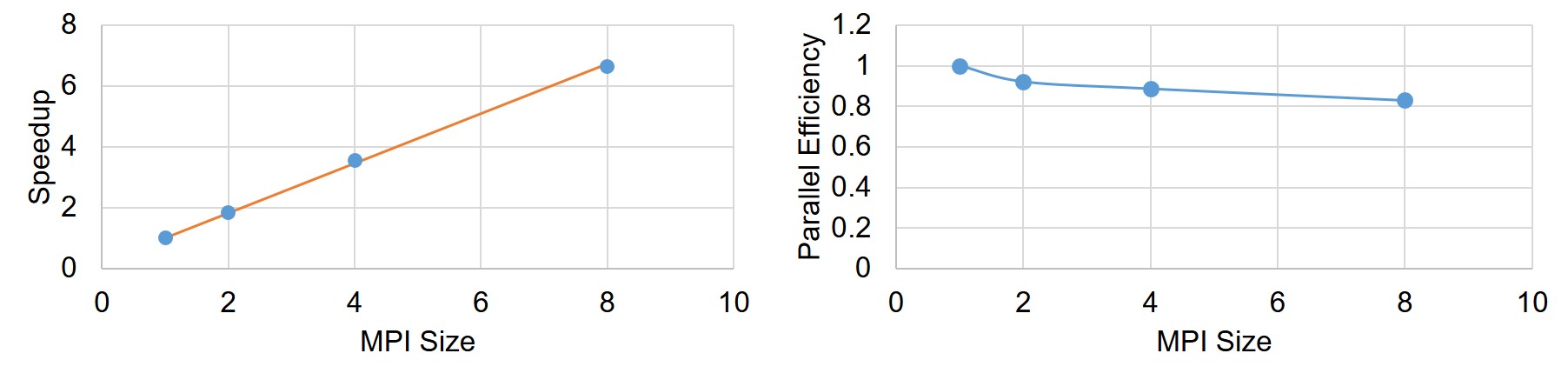}
\caption{Speed-up and parallel efficiency of the weak-scaling experiment. The orange line in the left plot shows the fitted curve based on the Gustafson equation.} \label{fig:weak_scaling_analysis}
\end{figure}

Speedup and parallel efficiency of the weak-scaling experiment is plotted in Fig. \ref{fig:weak_scaling_analysis}.  By fitting a curve based on the Gustafson equation (Eq. \ref{eq:weak_scaling}) on the obtained results (Fig. \ref{fig:weak_scaling_analysis}), the sequential proportion of the current implementation was calculated to be $18\%$, which means that  $82$ percent of the code can be parallelized, which is a proper but not an ideal scalability.

\begin{figure}[ht]
\center \includegraphics[width=14cm]{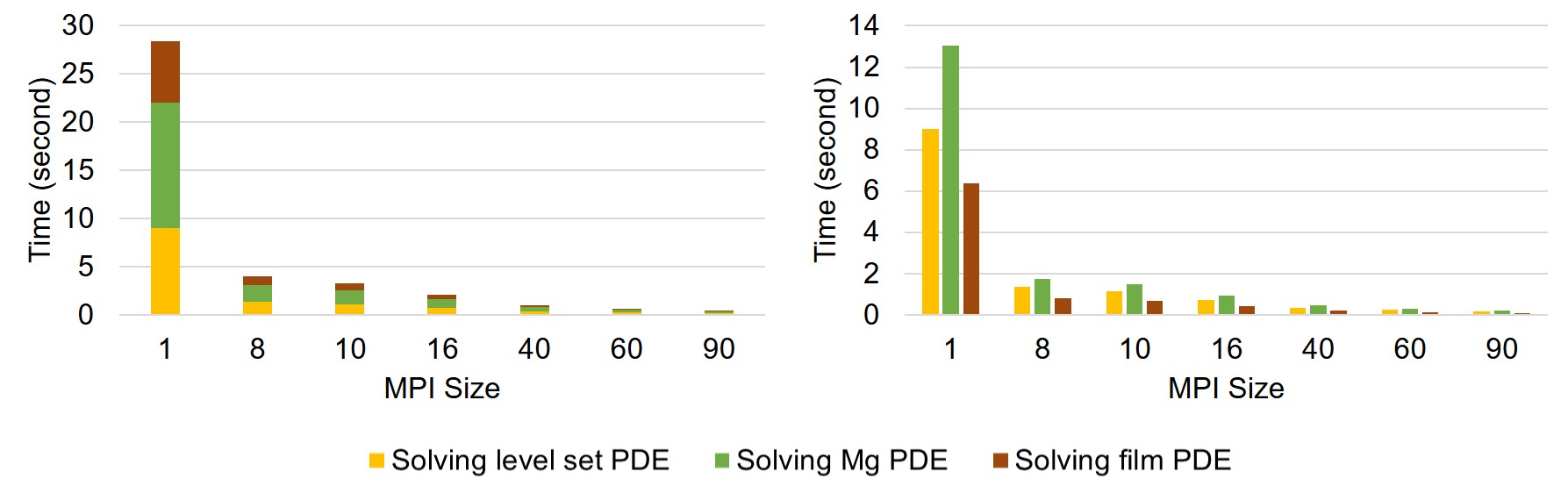}
\caption{Strong-scaling test result. Results are broken down into contributions  for each PDE, which are plotted  cumulatively and separately in the left and right plot, respectively.} \label{fig:strong_scaling_results}
\end{figure}

The strong-scaling results are plotted in Fig. \ref{fig:strong_scaling_results}, which shows a better scalability in comparison to the weak-scaling test. For a better representation, exact measured values are presented in Table \ref{tab:strong_scaling_results}.

\begin{table}[ht]
\caption{Strong-scaling test result, presented by the execution time of each PDE in simulations with a different number of employed MPI cores.}
\begin{tabular}{|c|l|c|c|c|c|c|c|c|}
\hline
\multicolumn{2}{|c|}{MPI Size}                                                                                     & 1     & 8    & 10   & 16   & 40   & 60   & 90   \\ \hline
\multirow{3}{*}{\begin{tabular}[c]{@{}c@{}}Solution time\\ of each time \\ step (s)\end{tabular}} & LS PDE & 9     & 1.39 & 1.14 & 0.75 & 0.36 & 0.26 & 0.19 \\ \cline{2-9} 
                                                                                                   & Mg PDE        & 13.04 & 1.76 & 1.48 & 0.94 & 0.46 & 0.31 & 0.22 \\ \cline{2-9} 
                                                                                                   & Film PDE      & 6.38  & 0.84 & 0.7  & 0.45 & 0.21 & 0.14 & 0.09 \\ \hline
\multicolumn{2}{|c|}{Total time (s)}                                                                               & 28.42 & 3.99 & 3.32 & 2.14 & 1.03 & 0.71 & 0.5  \\ \hline
\end{tabular}
\label{tab:strong_scaling_results}
\end{table}

\begin{figure}[ht]
\center \includegraphics[width=14cm]{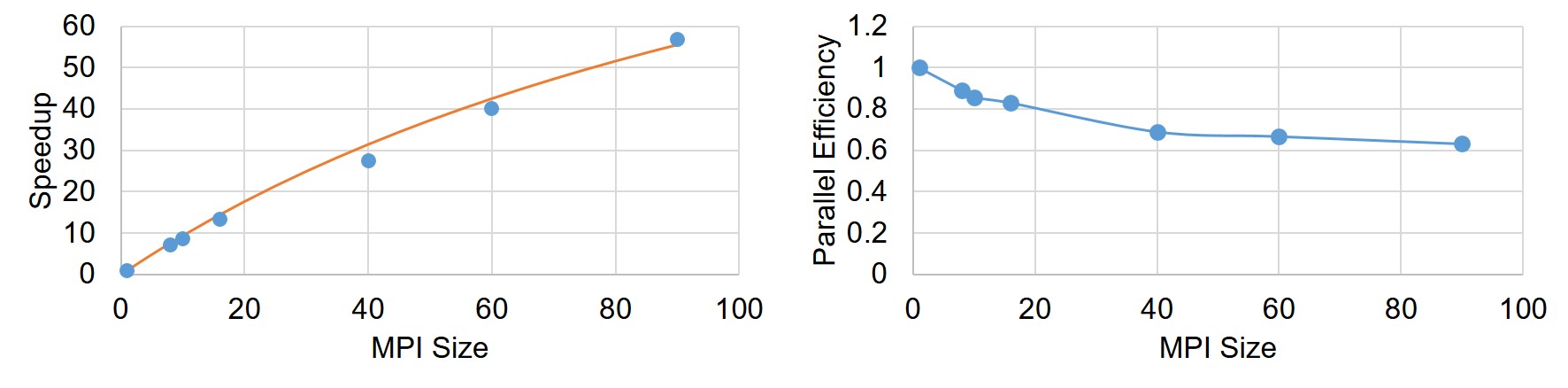}
\caption{Speed-up and parallel efficiency of the strong-scaling experiment. The orange line in the left plot is the fitted equation based on the Amdahl rule.} \label{fig:strong_scaling_analysis}
\end{figure}

Similar to weak-scaling results, Fig. \ref{fig:strong_scaling_analysis} demonstrates the speedup and parallel efficiency of the developed code for strong-scaling evaluation. From the results, it is obvious that increasing the number of cores leads to a better performance but a lower efficiency. By fitting Amdahl’s equation (Eq. \ref{eq:strong_scaling}) on the obtained speedup results (Fig. \ref{fig:strong_scaling_analysis}), $f$ was obtained as $0.01$, which means in strong-scaling terms that $99\%$ of the code can run in parallel. 

\section{Discussion}

In this investigation, the derivation and implementation of a reaction-diffusion model with moving boundaries were presented. Such an approach finds application in many scientific and engineering problems. The target application in the current work was the degradation of a bulk metal cuboid in a liquid environment, specifically Mg in an aqueous ion solution as a representative for temporary medical devices. The simulations were based on the corrosion of Mg metal to Mg ions to form a film of Mg hydroxide that partially protects the metal block from further degradation except where this film is impacted by reaction with other ions in the environment (such as chloride ion). The reactive moving boundary problem was cast in the form of equations in which the change of the concentrations of the different chemical components is represented by parabolic PDEs. The coupled equations depend on several kinetic constants that have been calibrated from experiments. The moving interface between the metal bulk and the liquid phase was described by an implicit function using the level set method. The derivation led to equations that require the use of numerical techniques for which a combination of finite difference and finite element methods was implemented. As the required high accuracy on the moving interface results in an increase in computation time, parallelization was crucial for the computational model to decrease the execution time of the simulations. The results of the total execution time in each time step (Table \ref{tab:strong_scaling_results}) clearly indicate that without the parallelization, the simulation of the model is slow and as a result, less interactable for real-world simulation analyses. Considering the properly employed parallelization, the computational time has been decreased noticeably for the investigated case-study.

The output of the conducted numerical simulation demonstrates that the developed mathematical model is capable of capturing the degradation interface movement and of modeling of the underlying chemical phenomena. The predicted mass loss is in line with the experimental results, and the simulated corrosion behavior is as expected for such a system. It is worth noting that the chosen system is highly idealized as a model for medical devices. A more realistic chemical environment would contain many more species that play a role in the formation of either soluble ions or the protective film. Moreover, in real-world scenarios, corrosion occurs in a more complex way than the simplified one described in this paper, which will have a significant influence on the local concentration of ions in the regions close to the solid surface. Nevertheless, the developed framework is capable of capturing these physical and chemical phenomena in the future by simply adding the appropriate terms to the base PDEs without any major changes in the computational model. Furthermore, although it requires some changes to the parallelization approach, the addition of the fluid flow around the block is feasible by adding convective terms to form a reaction-diffusion-convection system. Such a system can be used to model relevant systems such as experimental bioreactor setups in biology and medical sciences.

The parallel algorithm was implemented using a domain decomposition method. Standard domain decomposition preconditioners, such as restricted additive Schwarz, are widely used for parallel implementation of computational models. In a parallel implementation, such preconditioners bring the benefit of relatively low communication costs \cite{Daas2019AMS}. Beside this, the formed linear system of equations in each partition of the mesh was solved using Krylov methods by taking advantage of the highly-efficient preconditioners and iterative solvers of the PETSc library. According to the obtained results, the employed parallelization approach of the current study yields reasonable scaling with respect to the available computational resources (or the number of sub-domains). Out of multiple evaluations, the best performance was achieved using the preconditioner/solver combination of HYPRE/GMRES, which is in agreement with findings in more specific studies in this regard \cite{Ghai2018}.

To evaluate the scaling performance of the implemented parallelism, a set of weak and strong scaling tests was conducted. In weak-scaling, the main approach is changing the problem size proportional to the change in the available computing resources. In an ideal parallelization, we expect that the speedup remains the same for all the setups because we provide double resources as we double the size of the problem. In strong-scaling, the size of the problem remains constant, but the number of computing units increases. So, in an ideal case, we should observe a double speedup as the number of computing units doubles. 
By fitting Gustafson's and Amdahl's laws on the scaling test results (Figs. \ref{fig:weak_scaling_analysis} and \ref{fig:strong_scaling_analysis}), the maximum parallelizable portion of the code was calculated to be 82\% and 99\% for the weak-scaling and strong-scaling tests, respectively. This is a reasonable theoretical scaling for both cases. 

The obtained scaling behavior is similar to other conducted  studies for diffusion or diffusion-convection systems \cite{Hassan2011,Rettinger2017}, in which the efficiency of the parallelization decreases with increasing the number of available computational resources. The reason behind this behavior in the current model lies in the mesh partitioning process. Indeed, the mesh is partitioned into semi-equal partitions, each of which has the same number of elements, but the main computation is only carried out on the nodes located outside the degrading material block (i.e. in the medium). In other words, the computational resources assigned to the nodes inside the material bulk do not contribute significantly to the simulation. This limitation can be prevented by modifying the mesh generation process in a way that a lower number of elements be generated inside the material block, but doing this requires remeshing of the interior region as the moving interface approaches it, which imposes even more complexity to the algorithm due to the partitioned mesh. Another bottleneck of the current model, as discussed before, routed in the non-constant right-hand matrix of the linear system (Eq. \ref{eq:linear_system}), which requires computing the $A$ matrix (Eq. \ref{eq:system_a}) in each time step and leads to a slower execution time. 

One important point in this regard is that the way that the results are interpreted does not necessarily imply the true scaling behavior of the system. Indeed, it is more like a surrogate model of the system performance. The correct methodology for obtaining true scaling factors is rather starting from an analysis of the code and time used in each routine for a non-parallel run. Then, based on the fraction of routines that are possible to execute in parallel, one can get a theoretical limit for the speedup. This will be reduced by practical limitations such as load balancing and communication costs of the network. Since it is a theoretical limit, it is not fully correct to ignore those extra parts and use the execution time to invert the relation to predict the fraction of the code that is parallel. However, for a complex computational model like the one that was developed in the current study, doing such a measurement of each routine is very difficult due to the complexity of the orchestrated libraries and tools. As a result, we were limited to use the roughly approximated speedup limit to evaluate the scaling of the constructed model.

\section{Conclusion}

In this work, a mathematical model of a reaction-diffusion system with a moving front was constructed, and the corresponding computational model was implemented using the finite element method. In order to correlate the diffusion phenomenon to the moving boundary position, high numerical accuracy is necessary at the diffusion interface, which requires a finer discretization of space near the moving front. This leads to an expensive computational model, which makes employing HPC techniques crucial in order to improve the simulation execution time. To this end, a high-performance domain decomposition approach was employed to partition the mesh and distribute the workload to available computing resources. Additionally, an efficient preconditioner/solver combination for reaction-diffusion PDEs was used to optimize the model to be used for the high-performance simulation of large scale systems in which the movement of system boundaries is controlled by reaction-diffusion phenomena. 

The investigated problem was the degradation of a magnesium block inside a solution, in which the surface of the block moves due to the reaction-diffusion phenomena in the metal-medium interface. The implemented model showed a good agreement with the experimental data in terms of the degradation rate and chemical reactions, and the parallel efficiency and linear scalability were appropriate in performance evaluation tests. For the next stage of the study, it could be interesting to evaluate the model and its performance on a much larger system and tune the resources and memory usage by testing different preconditioners and solvers.

\section*{Acknowledgment}

This research is financially supported by the Prosperos project, funded by the Interreg VA Flanders – The Netherlands program, CCI grant no. 2014TC16RFCB046 and by the Fund for Scientific Research Flanders (FWO), grant G085018N . LG acknowledges support from the European Research Council under the European Union’s Horizon 2020 research and innovation programmen, ERC CoG 772418. The computational resources and services used in this work were provided by the VSC (Flemish Supercomputer Center), funded by the Research Foundation - Flanders (FWO) and the Flemish Government – department EWI.

\bibliographystyle{elsarticle-num}
\bibliography{mybibs}
\end{document}